\documentclass[10pt,twoside]{article}

\usepackage{8OSME-style}
\usepackage{mathcomp}

\title{Pillow Box Design}
\authorlist[Mitani]{J. Mitani} % the text in square brackets will be shown in the headers; this should be the author surnames
\affiliations{Jun Mitani \at University of Tsukuba, 1-1-1 Tennohdai Tsukuba Ibaraki, Japan, \email{mitani@cs.tsukuba.ac.jp} }

\makeatletter
  \let\runtitle\@title
  \let\runauthor\shortauthor
\makeatother

\begin{document}

\maketitle

\begin{abstract}
This paper focuses on packaging design using origami techniques, specifically designs incorporating curves, known as pillow boxes. While conventional paper packaging boxes are typically cuboid, pillow box designs include curved surfaces, offering both aesthetic and practical advantages. This study analyzes the specific curved folds of pillow boxes, clarifying the fundamental geometric condition these curves must meet. Additionally, it proposes new design variations for pillow boxes based on the condition. The relationship between the shape of the folds and the volume of the final three-dimensional shape is also explored. This research extends the boundaries of functionality and aesthetics in origami design and explores new possibilities in packaging solutions.
\end{abstract}

\section{Introduction}
\label{sec:introduction}
One of the notable applications of origami techniques in practical use can be observed in the design of packaging boxes. Boxes made from paper come in many variations, ranging from simplistic designs to those that are intricately crafted. Among these, some even have curved folds. 
The {\it pillow box} can be cited as one of the most simplistic yet popular packaging designs that incorporate curved folds. 

Figure 1a shows an example of a pillow box, while Figure 1b presents its crease pattern, i.e. unfolded layout. 
Curved creases can also be observed in the French fry containers used at hamburger chain restaurants. The pillow box can be considered as a form composed of two of these containers joined together.
Different three-dimensional shapes of boxes can be made depending on the shape of the curved folds of the pillow box shown in Figure 1b.
However, there are specific conditions that these curves must fulfill for a design to function as a box.
This paper aims to clarify the geometrical constraint of the curves used for pillow boxes, with the objective of expanding the design variations. Additionally, the latter part of the paper explores the relationship between the shape of the folds and the volume of the box.

The crease pattern of the pillow box targeted in this paper, as depicted in Figure 1b, exhibits symmetry both vertically and horizontally. After folding, the three-dimensional shape possesses symmetry along the z-axis and y-axis directions. Upon folding, the sides overlap, resulting in a two-layered structure. 

As shown in Figure 1c (a quarter of the three-dimensional shape), the shape after folding consists of both the top and side surfaces being cylindrical. The rulings that constitute the top surface are horizontal, while those forming the side surface are vertical. By mirror-inverting a part of the top surface over a plane inclined at 45°, the side surface can be formed. This means that the three-dimensional folds of the pillow box are planar curves.

\begin{figure}[htb]
  \centering
  \includegraphics[width=0.9\textwidth]{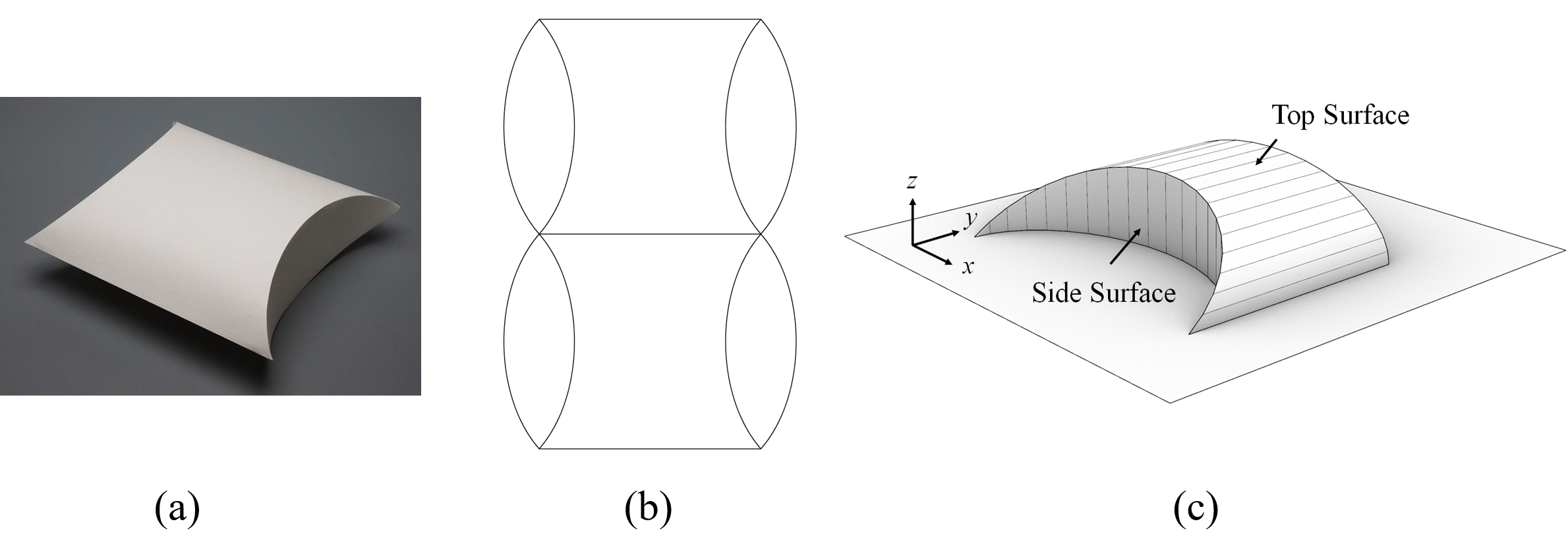}
  \caption{Photo of a pillow box (a), its crease pattern (b) and one-fourth of the entire three-dimensional shape.}
  \label{fig01}
\end{figure}

\section{Related Work}

\subsection{Curved origami}

The curved surfaces created by bending paper are limited to a specific type of ruled surface, where a straight line moves through space, characterized by a Gaussian curvature of zero; these are known as developable surfaces. Non-planar developable surfaces are confined to conical, cylindrical, and tangential surfaces. However, by incorporating folds into these surfaces, it is possible to create a wide variety of shapes.

Many studies have been conducted on the geometry of developable surfaces with curved folds. Fuchs and Tabachnikov \cite{Tabachnikov1999} have clarified the relationship between three-dimensional spatial curves that become folds, their corresponding curves mapped onto a crease pattern, and the angles of the folds. Tachi streamlined this into clear equations \cite{Tachi2011}, paving the way for shape design with curved folds. Demaine et al. have explored the geometric properties of developable surfaces with curved folds, uncovering many facts \cite{demaine2015}.
When the fold is a planar curve, it is characterized by equal folding angles on both sides of the fold. This can be seen as dividing the developable surface into two parts with a plane and then mirror-inverting one side. This principle has been widely used in the creation of origami art by David Huffman \cite{demaine2010}. Utilizing this principle of mirror inversion, Mitani and Igarashi developed a tool for interactively designing shapes that include planer curved folds \cite{Mitani2011}.

\subsection{Volume of the pillow box}
The volume of the pillow box is discussed in the latter part of this paper. 
Related to this is a discussion that took place in 1997-1998 on a web page named The Geometry Junkyard (which remains in the web archive \cite{web1}), in which various contributors discussed the geometrical problem of maximising the volume inside two square sheets of length one when sealed along their edges. 
At the end of the web page compiled by Andrew Kepert on these entitled 'Teabag problem' (which also remains in the web archive \cite{web2}), there is a description of the shape of the pillow box discussed in this paper, where the maximum volume is approximately 0.1703844172 when $\theta$ is approximately 1.047 radians under the condition that the surface is along an arc of angle $2 \theta$ (the value we have obtained with our optimisation method is larger than this.)

\section{Variation of Shapes}
\label{sec:section}

\subsection{Symmetric Shape with Respect to the Horizontal Plane}

The shape targeted in this paper is discretized into a collection of elongated planar rectangles, as illustrated in Figure 2a.

\begin{figure}[htb]
  \centering
  \includegraphics[width=.9\textwidth]{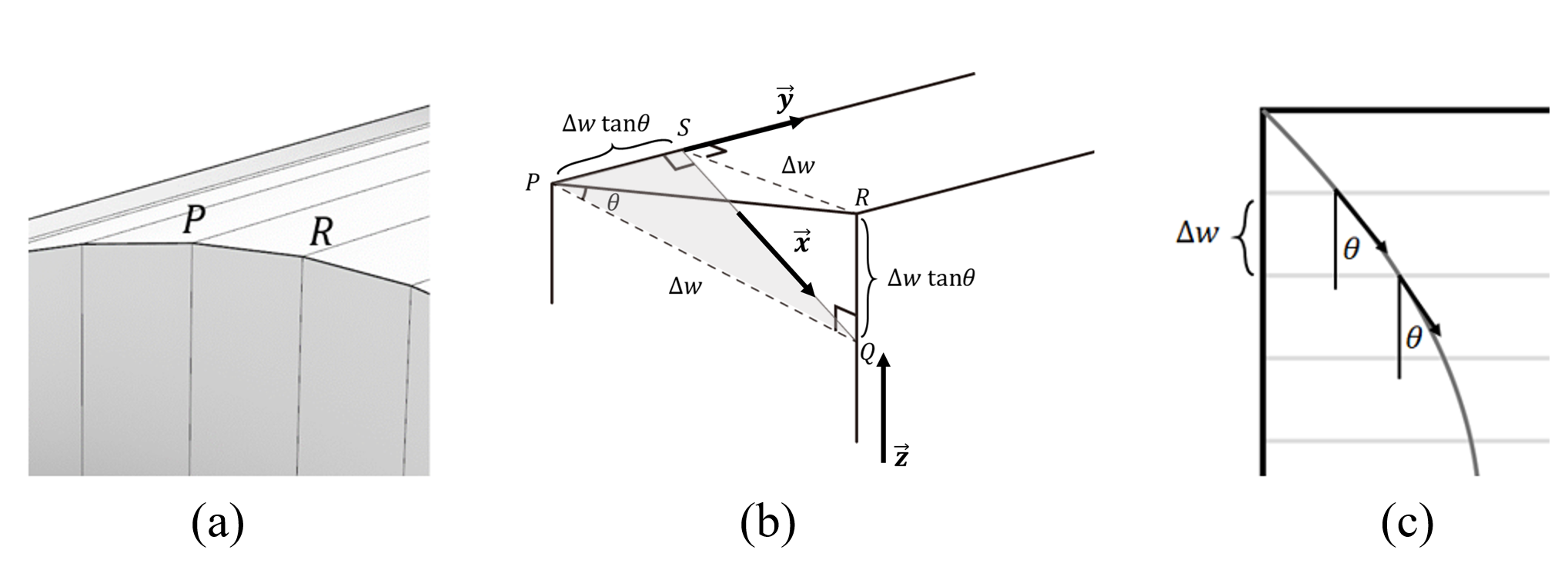}
  \caption{Discrete model of the pillow box.} %% ラベルを付ける
  \label{fig02}
\end{figure}

Due to the geometric constraint that the area of $\triangle$SPQ in Figure 2b must be non-negative, the angle $\theta$ needs to fall within the range specified by the following equation.

\begin{equation}
    - \pi / 4 \leq \theta \leq  \pi / 4
    \label{eq:numbered}
\end{equation}

This fact is explained as follows:

Figure 2b is an enlargement of the planar rectangles on both sides of edge PR from Figure 2a. At points P and R, edges are connected that are parallel to the $z$-axis and $y$-axis. The distance $w$ is the distance between the vertical lines passing through P and R. Q is the foot of the perpendicular dropped from P to the vertical edge passing through R. S is the foot of the perpendicular dropped from R to the horizontal edge passing through P. If $\angle RPQ$ is denoted as $\theta$, then PS and RQ are equal to ${\Delta}w \tan \theta$. Since $\triangle$SPQ is a right triangle lying horizontally, the following equation holds based on the Pythagorean theorem:

\begin{equation}
\begin{aligned}
    PS^2 + SQ^2 &= PQ^2 \\
    ({\Delta}w \tan{\theta})^2+SQ^2 &={\Delta}w^2 \\
    ( \tan{\theta} )^2 &= 1 - \frac{SQ^2}{{\Delta}w^2} 
    \label{eq:numbered1}
\end{aligned}
\end{equation}

Therefore
\begin{equation}
    | \tan{\theta} | \leq 1 \\
\end{equation}
and the following equation holds    
\begin{equation}
    - \pi / 4  \leq \theta \leq \pi / 4
\end{equation}

\begin{figure}[htb]
  \centering
  \includegraphics[width=.5\textwidth]{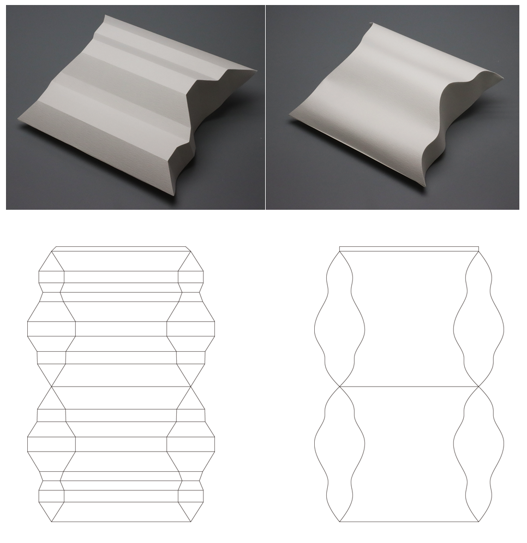}
  \caption{Out-of-the-ordinary pillow box design. Polyhedral model and curved surface model.}
  \label{fig03}
\end{figure}

The value of $\theta$ corresponds to the angle of the tangent to the fold curve in the crease pattern shown on the right side of Figure 2. Consequently, the constraint for the curve of the fold is that the absolute value of the angle of its tangent must be 45 degrees or less.

Under this condition, when fold curves are created on the crease pattern, a pillow box can be made by folding them. Figure 3 shows two examples of pillow boxes created with designed folds while satisfying the constraints. The left example features a polyline fold, while the right one is designed with smooth curved folds.

The constraint on the shape of the curve is simple, and under this constraint, it was possible to create various shapes of pillow boxes as shown in Figure 4.

\begin{figure}[htb]
  \centering
  \includegraphics[width=1.0\textwidth]{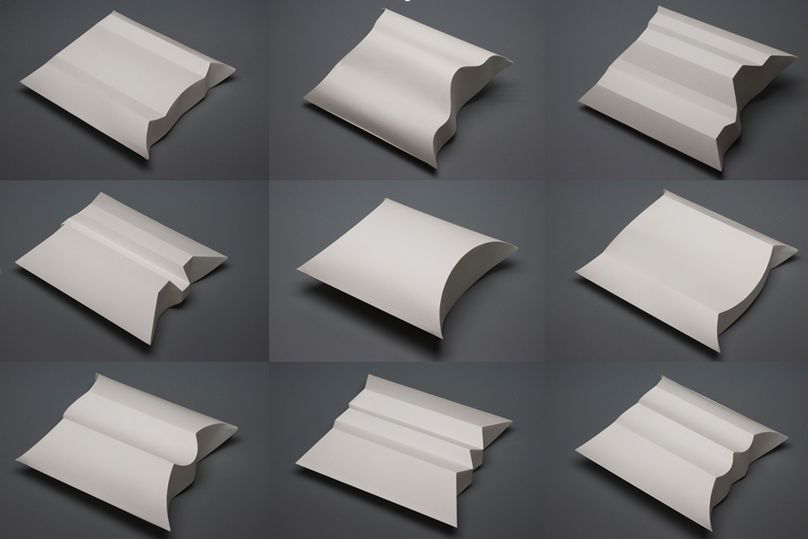}
  \caption{Various pillow box designs.}
  \label{fig04}
\end{figure}

\subsection{Asymmetric Shape with Respect to the Horizontal Plane}
\label{subsec:subsection}
In previous designs of pillow boxes, a constraint was imposed such that their shape was symmetrical with respect to the $xy$-plane, meaning that the rulings composing the side surface were parallel to the $z$-axis. This section demonstrates that by relaxing this constraint, it is possible to increase the variety of designs further.

\begin{figure}[htb]
  \centering
  \includegraphics[width=.6\textwidth]{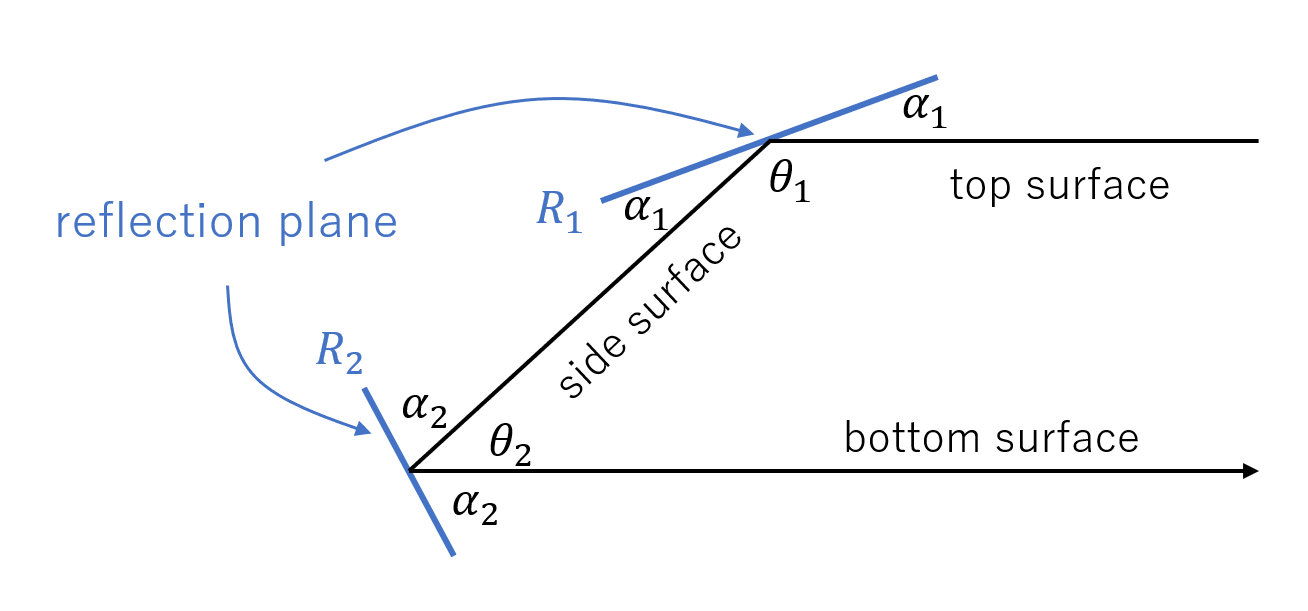}
  \caption{Cross-sectional view of a pillow vox parallel to the $yz$-plane, where rulings appear reflected by reflection planes}
  \label{fig05}
\end{figure}

Figure~\ref{fig05} presents a sectional view of a pillow box parallel to the $yz$-plane. As previously mentioned, the rulings that constitute the top surface, side surface, and bottom surface are all parallel to the $yz$-plane. Therefore, in this sectional view, the reflection planes $R_1$ and $R_2$ appear to reflect the rulings.

Assuming both the top and bottom surfaces are horizontal, the following relationships hold for the angles $\alpha_1$, $\alpha_2$, $\theta_1$, $\theta_2$ depicted in the figure:

\begin{equation}
\theta_1 + \theta_2 = 180\tcdegree, \, \alpha_1 = \frac{\theta_2}{2}, \, \alpha_2 = \frac{\theta_1}{2}.\\
\end{equation}
Thus, determining one of these values automatically determines the remaining values.

In the designs discussed in the previous section, both $\alpha_1$ and $\alpha_2$ were set at 45 degrees, and both $\theta_1$ and $\theta_2$ were 90 degrees. However, setting different values allows for the creation of pillow box shapes where the side surface is not perpendicular to the $xy$-plane.

\begin{figure}[htb]
  \centering
  \includegraphics[width=.7\textwidth]{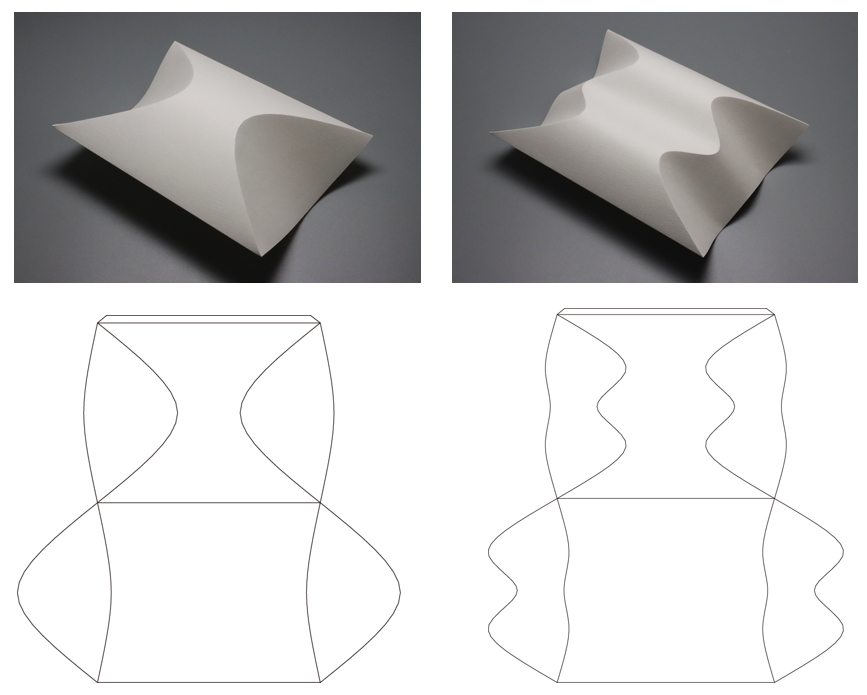}
  \caption{Designs of pillow boxes not symmetrical with respect to the $xy$-plane.}
  \label{fig06}
\end{figure}

In the examples shown in Figures~\ref{fig03} and \ref{fig04}, curved creases were drawn on the development and then physically folded. However, calculating the shape of curves on the development using the approach in this section is challenging. Therefore, by initially creating a cylindrical top surface and applying two mirror transformations corresponding to the reflections at $R_1$ and $R_2$ in Figure~\ref{fig05}, it is possible to design pillow boxes that are not symmetrical with respect to the $xy$-plane as shown in Figure~\ref{fig06}.

\section{Relationship between Curve Shape and Volume}

As seen in the previous section, the shape of the folds in a pillow box affects its form, and consequently, its volume also changes. An intriguing question arises: what is the shape of the fold curve that maximizes this volume, and what is the volume in that case? In the followings, this question will be discussed in detail. This exploration delves into the relationship between the design of the crease pattern and the functional capacity of the pillow box.

\subsection{Preliminary}

Here we revisit and organize the conditions for the development of the pillow box we are focusing on.

\begin{figure}
  \centering
  \includegraphics[width=.8\textwidth]{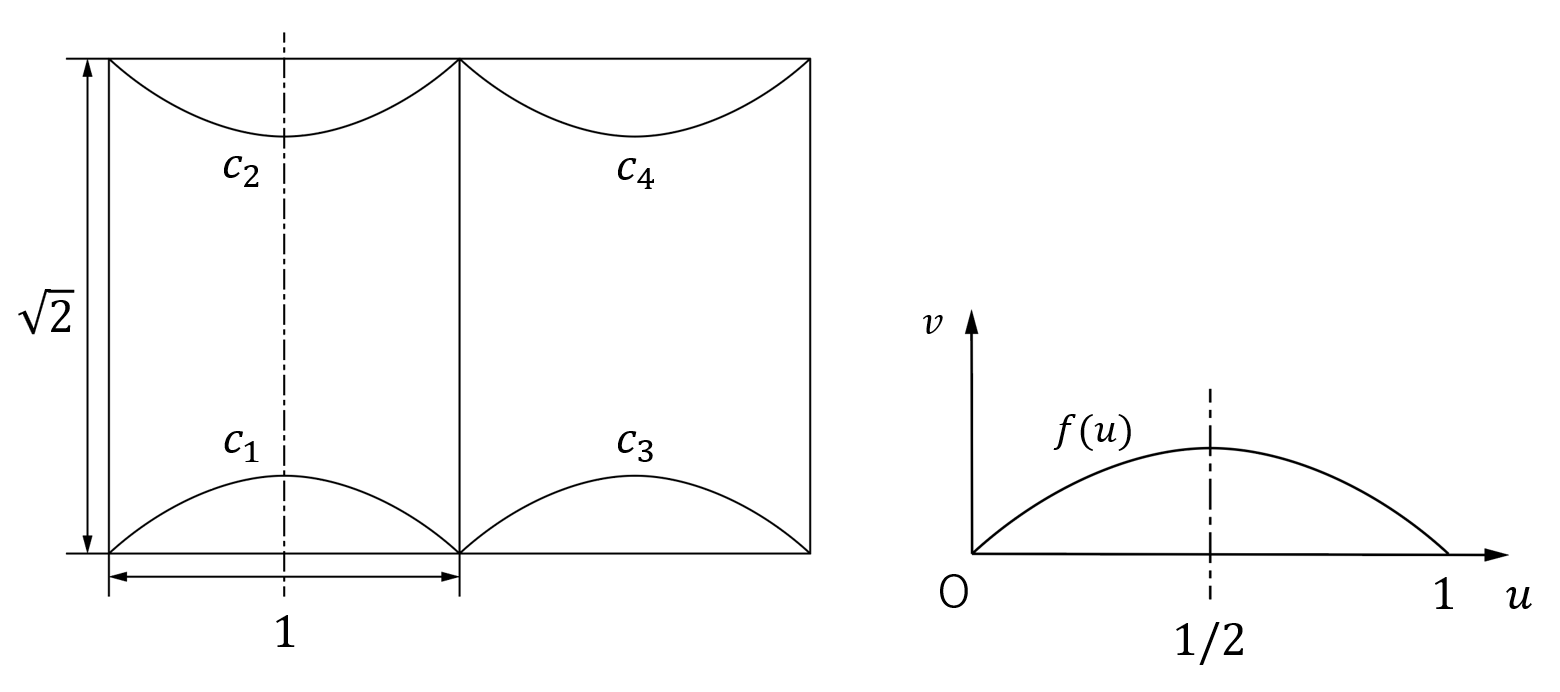}
  \caption{Curves serving as creases in a pillow box.}
  \label{fig07}
\end{figure}

Firstly, whereas previously it was assumed that the side surfaces overlapped in two layers, we will now consider the side surface to consist of a single layer. Consequently, the development becomes a rectangle, as shown in the left side of Figure~\ref{fig07} (which is positioned as if rotated 90 degrees from the development in Figure 1). This implies that a pillow box can also be created by adding curved creases to a generally rectangular envelope.

In the discussions that follow, we will assume that the curves $c_1$ to $c_4$ defining the shape of the pillow box are all of the same shape. Additionally, each curve is symmetrical left to right (with curve $c_1$ being symmetrical about the centerline of the figure). It can be easily understood from the following discussion that to find a curve that maximizes volume, it suffices to examine curves that are symmetric left to right; If the shape of the curve is not symmetrical, by comparing the volumes of the left and right sides of the solid and selecting the side that does not have a smaller volume, then adjusting the shape of the curve on the opposite side to mirror the selected side, the volume obtained after this operation will be equal to or greater than the volume of the original shape. Therefore, it is impossible for an asymmetrical shape to yield a unique maximum volume.

Furthermore, let's assume as a premise that the size of the original envelope is a rectangle with a width of 1 and a height of $\sqrt{2}$. It's important to note that if this ratio is different, the shape of the curve that maximizes the volume will also change. For example, if the height is infinitely large, the volume is maximized when the cross-sectional shape is a circle. 

The fold curve on the development will be represented by the function $f(u)$ when plotted in an orthogonal $uv$-coordinate system, as depicted in the right side of Figure~\ref{fig07}. This function $f(u)$ is symmetrical with respect to the line $u=\frac{1}{2}$ and satisfies $f(0)=f(1)=0$. Additionally, based on the condition derived in the previous section, the function must satisfy the condition $|f'(u)| \leq 1$.

In the followings, we will consider examples where the cross-section in the $xz$-plane is a circle, a rectangle, or a rhombus, as these shapes allow for a simpler calculation of the volume. After that, we will consider the shape with an arched cross-section commonly seen in pillboxes.

\subsection{Circular cross section}

When the cross-section is a circle, given that the circumference of the circle is 2, the radius $r$ is $\frac{1}{\pi}$. 
The function $f(u)$ is expressed as follows:
\begin{equation}
f(u) = \frac{\sin (\pi u)}{\pi} 
\end{equation}

\begin{figure}[htb]
  \centering
  \includegraphics[width=.8\textwidth]{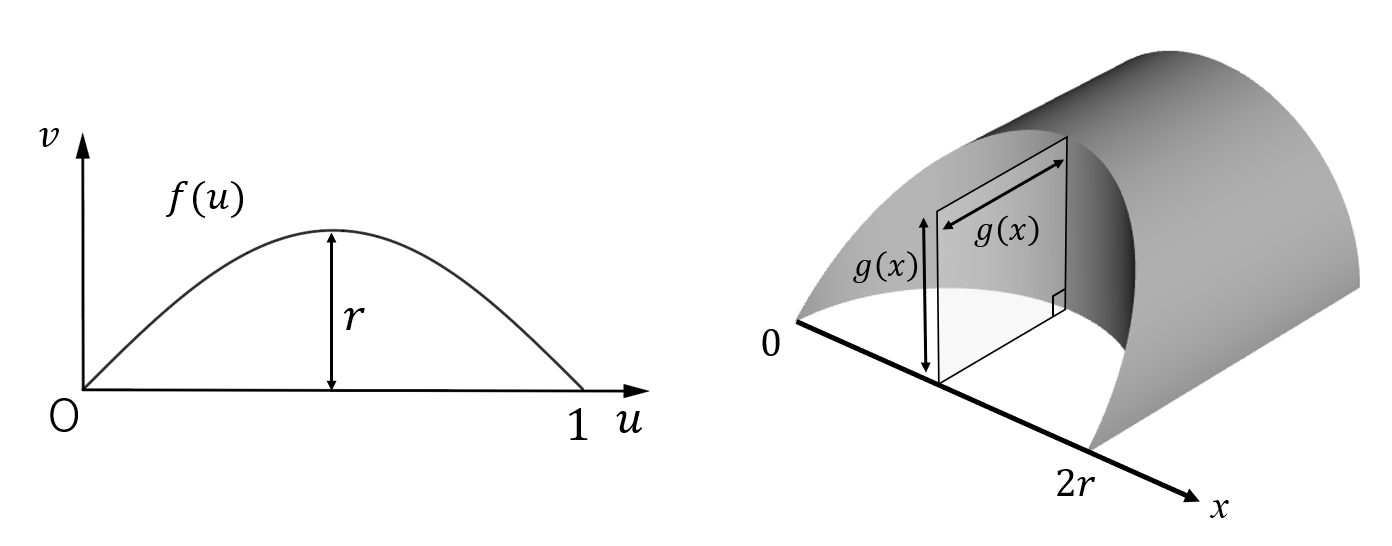}
  \caption{The shape of one quarter of a pillow box with a circular cross-section.}
  \label{fig08}
\end{figure}

The graph in the left side of Figure~\ref{fig08} represents the fold curve and the figure in the right side represents quarter of the resulting three-dimensional shape. 

The volume of this shape can be calculated as a quarter of a cylinder with a radius of $\frac{1}{\pi}$ and a length of $ \sqrt{2}$, from which the volumes of a part
\begin{equation}
\int_{0}^{2r} \left(g(x)\right)^2 \ dx
\end{equation}
is subtracted. 
Since \(g(x)\) represents a semicircle centered at \(x = r\) with radius \(r\), it can be expressed as follows:

\begin{equation}
g(x) = \sqrt{r^2 - (r-x)^2} = \sqrt{2rx-x^2}
\end{equation}
Therefore, the volume of a pillow box with a circular cross section can be calculated as follows. 
\begin{equation}
\sqrt{2}r^2\pi-4\int_{0}^{2r} \left(g(x)\right)^2 \ dx \approx 0.27815
\end{equation}

\subsection{Rectangular cross section}

\begin{figure}[tb]
  \centering
  \includegraphics[width=.8\textwidth]{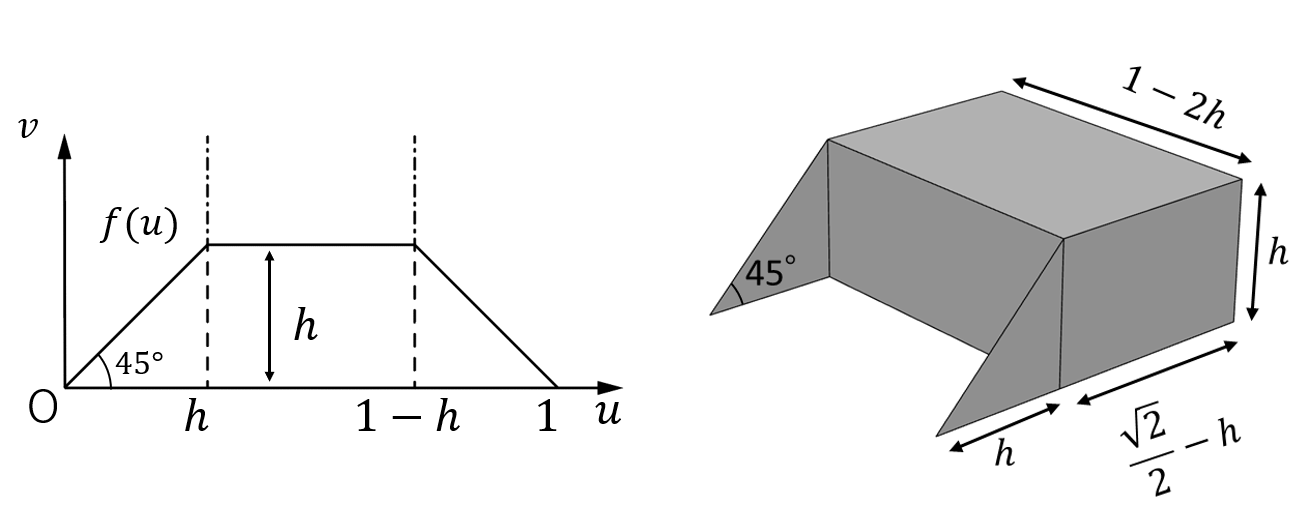}
  \caption{The shape of one quarter of a pillow box with a rectangular cross-section.}
  \label{fig09}
\end{figure}

Assuming the cross-section is rectangular, by adding fold lines at a 45-degree angle from both endpoints as shown in Figure~\ref{fig09}, a rectangular box shape can be created in the manner of a caramel box. As depicted in the figure, if the distance from the endpoints to the rectangular prism is denoted as \(h\), then the height, width, and depth of the rectangular prism are \(2h\), \(1-2h\), and \(\sqrt{2} - 2h\) respectively, thus the volume can be expressed by the following equation.

\begin{equation}
2h (1 - 2h) (\sqrt{2} - 2h)
\end{equation}

This expression reaches its maximum value 
$$1/27(4-\sqrt{2}-2\sqrt{6-2\sqrt{2}}+6\sqrt{3-\sqrt{2}}) \approx 0.24369$$
when 
$$h=\frac{1}{6}+\frac{1}{3\sqrt{2}}-\frac{\sqrt{3-\sqrt{2}}}{6}.$$

\subsection{Rhombic cross section}

\begin{figure}[tb]
  \centering
  \includegraphics[width=.8\textwidth]{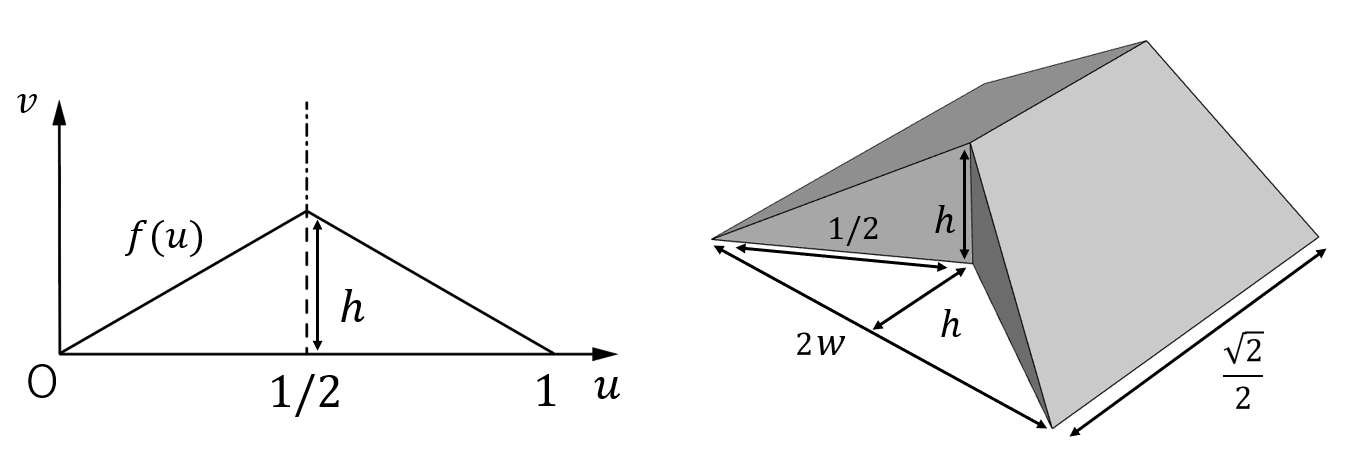}
  \caption{The shape of one quarter of a pillow box with a rhombic cross-section.}
  \label{fig10}
\end{figure}

If the cross section is a rhombus as shown in Figure~\ref{fig10}, the volume can be expressed by the following equation, where $h$ is the height of the triangle formed by the fold line and the $u$-axis, 

$$4( \frac{1}{2}\sqrt{2}wh-\frac{2}{3}h^2w) $$

where 

$$w=\sqrt{\frac{1}{4}-h^2}.$$

Since the maximum volume could not be analytically determined even by using symbolic computation software, it was confirmed that the maximum volume, approximately 0.24351, occurs when the value of $h$ is approximately 0.30547 by a numerical approach. 

\subsection{Arched cross section}

\begin{figure}[tb]
  \centering
  \includegraphics[width=.8\textwidth]{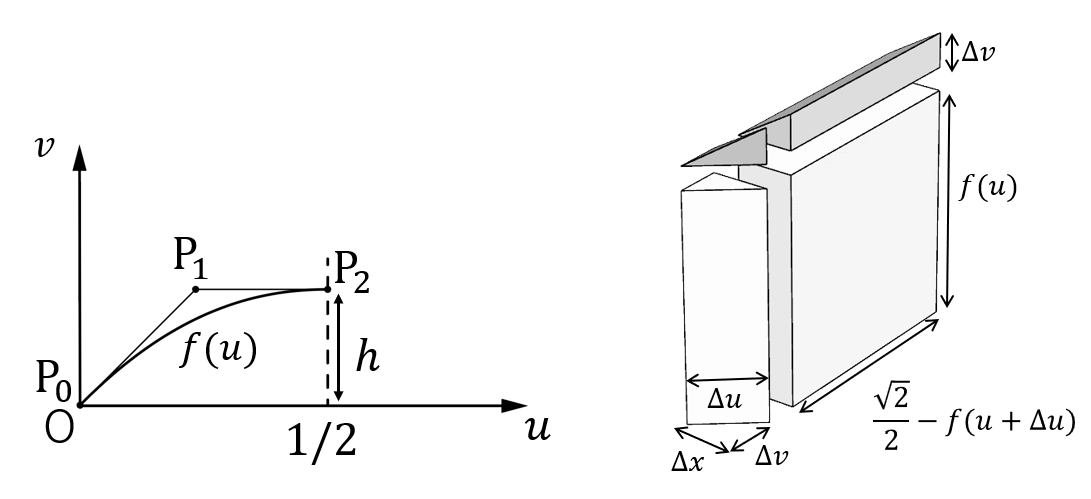}
  \caption{Crease curve defined by Bézier curve and discrete model.}
  \label{fig11}
\end{figure}

To address a more general arch-shaped top surface, here the segment of the crease curve within the range \(u \in [0,1/2]\) is represented by a quadratic Bézier curve. Given that the crease curve is symmetric, the segment for \(u \in [1/2, 1]\) is the reflection of this quadratic Bézier curve about \(u=1/2\). The shape of this Bézier curve is determined by the positions of control points $P_0$, $P_1$, and $P_2$, as depicted in Figure~\ref{fig11}. Since the endpoints of the crease curve coincide with the origin of the $uv$-coordinates, the position of $P_0$ is fixed at the origin. Additionally, the $u$-coordinate of $P_2$ is $1/2$.

%Additionally, let's assume the angle formed by $P_0P_1$ is 45 degrees and $P_1P_2$ is horizontal. This configuration is considered reasonable for maximizing volume. 
Under this setup, the coordinates of each control point can be represented as follows, with the $v$-coordinate value of $P_2$ being denoted as \(h\):

\begin{equation}
P_0(0,0), P_1(a, b), P_2(\frac{1}{2},h) 
\end{equation}

In essence, the shape of the curve is determined by three variable $a$, $b$, and \(h\).
In this case, the B\'ezier curve is expressed by the following equation using the parameter $t \in [0,1]$:

\begin{equation}
\begin{cases}
    u(t) = 2a(1-t)t+\frac{1}{2}t^2 \\
    v(t) = 2b(1-t)t+ht^2
\end{cases}
\label{eq:two-align-one-number}  
\end{equation}

%This setup defines the B\'ezier curve for the fold, providing a framework for designing a variety of arch-shaped top surfaces on pillow boxes.

The three-dimensional shape obtained from folding along this curve is discretized, as shown in the right side of Figure~\ref{fig11}. By calculating the sum of the volumes of each discrete element, the total volume of the shape can be determined. When the discretization interval \( \Delta t \) is set to 0.001, it was found that the maximum volume, approximately 0.294436, is achieved when $a=b=0.2731$, and $c=0.2544$.
To obtain these values, the SLSQP algorithm, which is widely used to solve constrained optimisation problems, was used. The optimise function \texttt{optimize. minimize} from the \texttt{scipy} library for Python was used for this purpose.

\begin{figure}[htb]
  \centering
  \includegraphics[width=.9\textwidth]{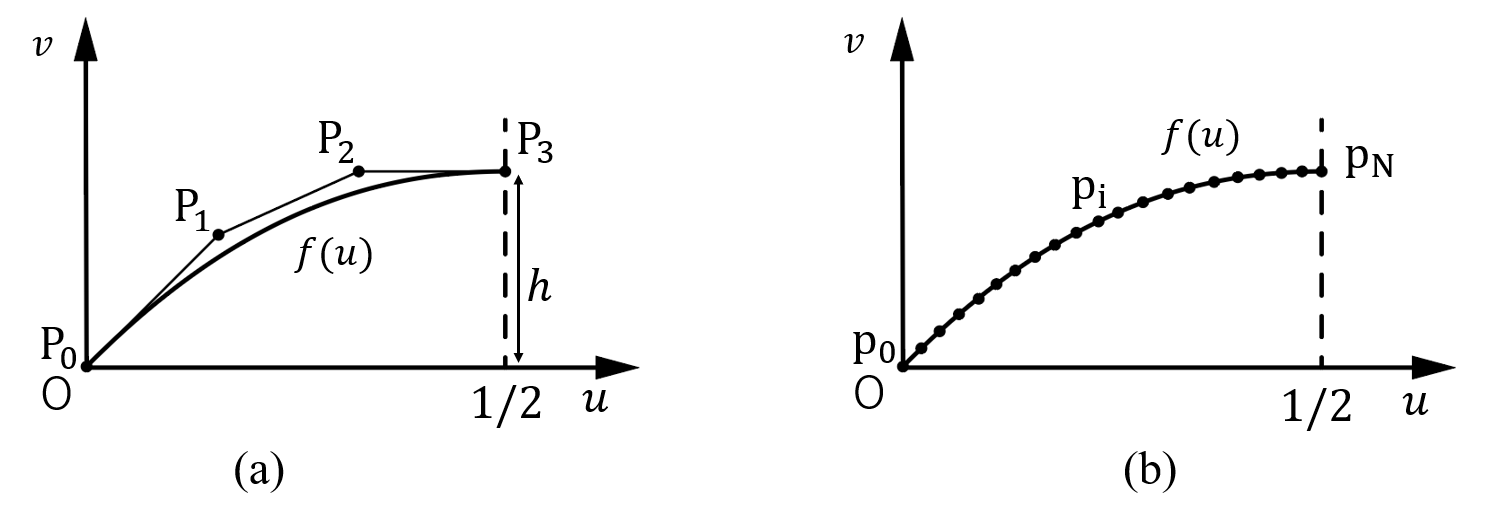}
  \caption{Crease curve defined by a cubic-Bézier curve and a fine polyline.}
  \label{fig12}
\end{figure}

Furthermore, as an attempt to achieve larger values, experiments were conducted to optimize the crease curve using two representations: a cubic Bézier curve (as shown in Figure~\ref{fig12}(a)) and a fine polyline (as shown in Figure~\ref{fig12}(b)).

For the cubic Bézier curve, the shape is defined by four control points from $P_0$ to $P_3$ using five variables: \(a\), \(b\), \(c\), \(d\), and \(h\), each represented by the following coordinate values:
\begin{equation}
P_0(0,0), P_1(a, b), P_2(c, d), P_3(\frac{1}{2},h). 
\end{equation}

The curve is expressed by the following equation using the parameter $t \in [0,1]$:

\begin{equation}
\begin{cases}
	u(t) = 3(1-t)^2ta + 3(1-t)t^2c + \frac{1}{2}t^3\\
	v(t) = 3(1-t)^2tb + 3(1-t)t^2d + t^3h
\end{cases}
\label{eq:two-align-one-number1}  
\end{equation}

By discretizing this curve in the same manner as before and calculating the volume, and solving it as an optimization problem with five variables, the maximum value of 0.295448 was obtained when $a=b=0.1125$, $c=d=0.2526$, and $h=0.2543$.

For the fine polyline, a polyline with 1000 segments was used (setting the $N$ value in Figure~\ref{fig12}(b) to 1000). Starting from the origin as the 0-th point, the $u$-coordinate of the $i$-th point \(p_i\) was set as \(i/2000\), and the $v$-coordinate was determined through optimization. This approach treated the problem as an optimization with 1000 variables. As a result, a value of 0.295449 was obtained for the shape shown in Figure~\ref{fig13}.

\subsection{Summary of evaluation}

Throughout this study, the maximum volumes for various cross-sectional shapes were determined. The results are summarized in Table 1. Among the easily creatable shapes—circle, rectangle, and diamond—the circular cross-section was found to have the largest volume. 

For the general pillow box shape with an arch-shaped top surface, using a cubic Bézier curve yielded a maximum volume approximately 0.3\% larger than when using a quadratic Bézier curve. 
Furthermore, employing a 1000-segment polyline allowed for a slight increase in volume, but the difference compared to using a cubic Bézier curve was extremely small, at only 0.0003\%. 

These findings suggest that the values obtained here are likely very close to the upper limit of the volume that can be realized with a pillow box made from an envelope of width 1 and length \(\sqrt{2}\).

Using this method, the maximum volume value was also determined when the sheet was a square with a length of one side. This was 0.174628, as shown in the table. This value is 2.49\% greater than the value of 0.1703844172 calculated by Andrew Kepert as presented in Section 2.2.

\begin{figure}[tb]
  \centering
  \begin{minipage}{0.49\textwidth}
    \centering
    \includegraphics[width=\textwidth]{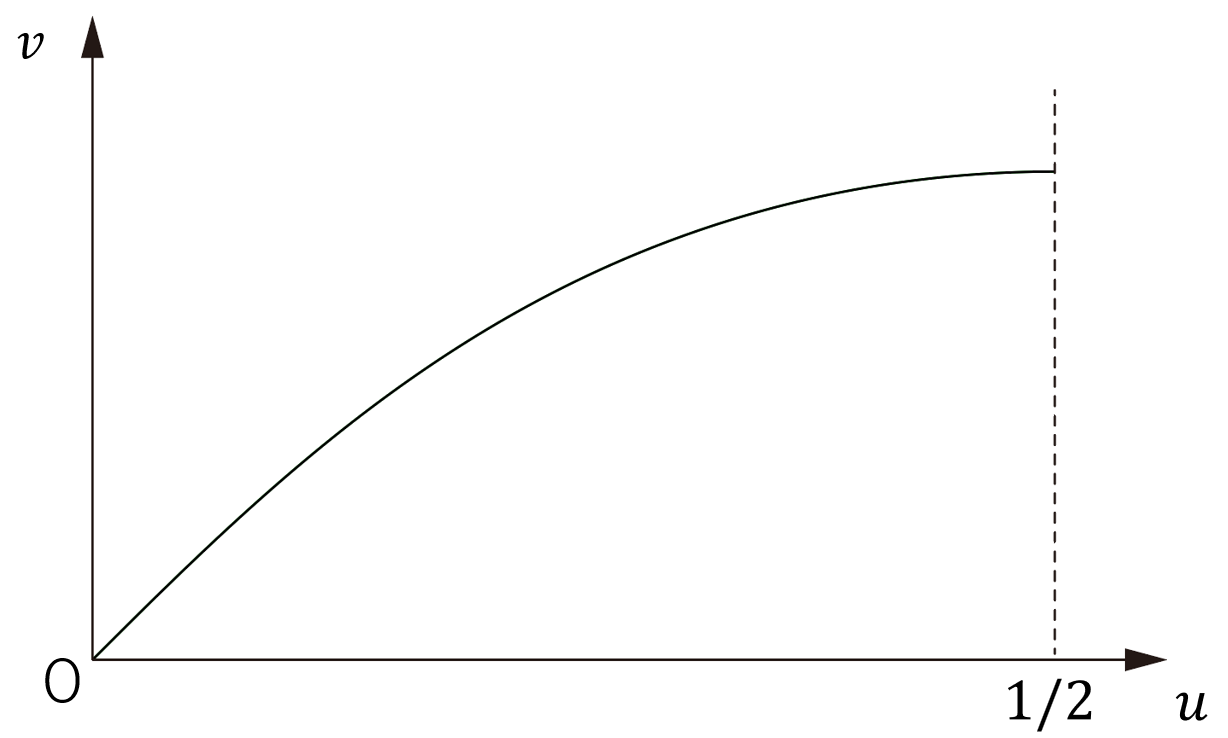}
    % オプションでここにキャプションを付けることができます。
  \end{minipage}\hfill
  \begin{minipage}{0.49\textwidth}
    \centering
    \includegraphics[width=\textwidth]{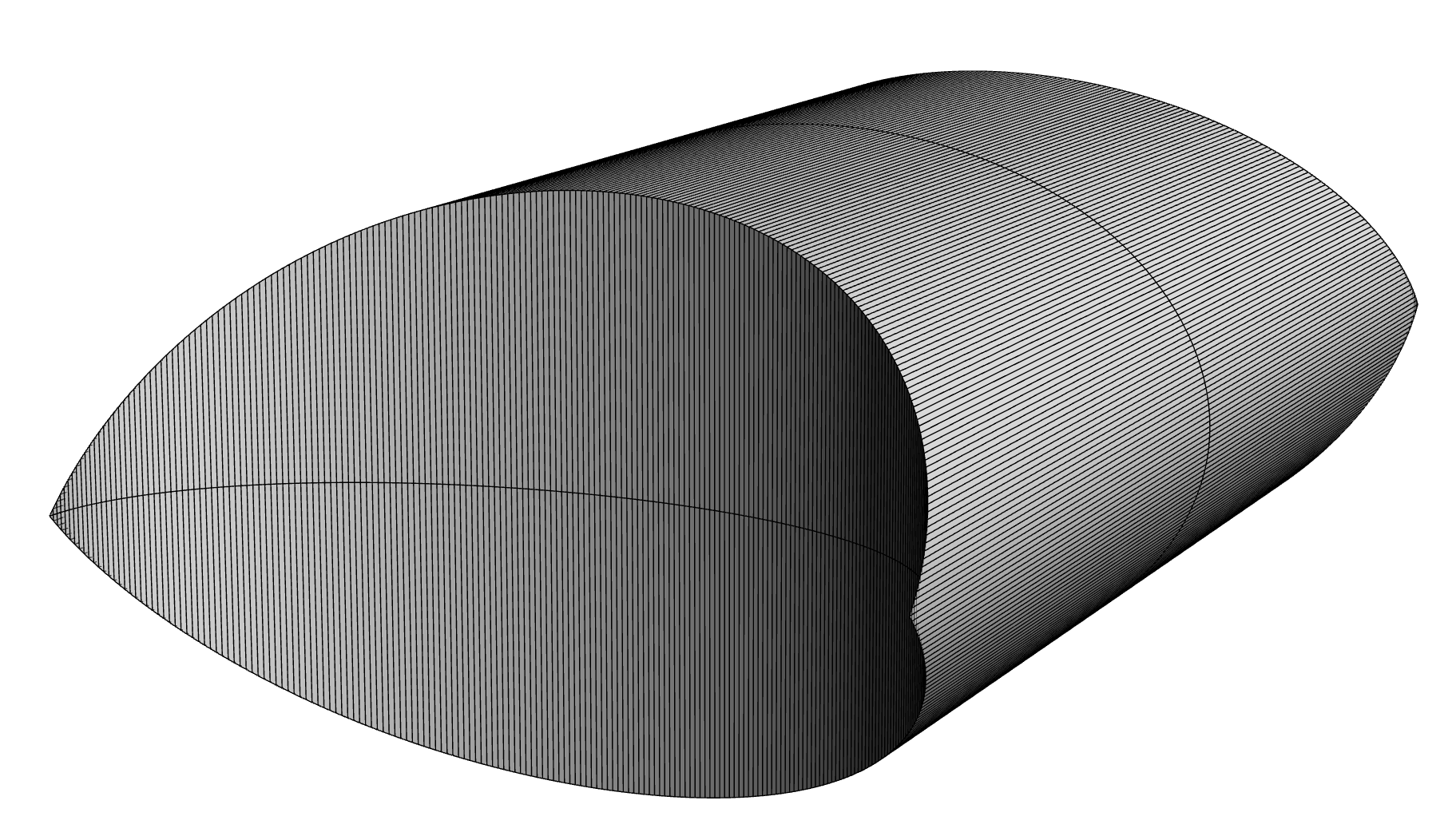}
    % オプションでここにキャプションを付けることができます。
  \end{minipage}
  \caption{Volume-maximising curve (1000-segment polyline) and the 3d model (100-segment polyline).}
  \label{fig13}
\end{figure}

Figure~\ref{fig13} depicts the 1000-segment polyline that maximizes volume. The cubic Bézier curve that maximizes volume is almost identical in shape to this polyline, to the extent that if drawn on the same graph, the lines would overlap and appear as a single line, making it impossible to distinguish between them.

\begin{table}[htb]
  \centering 
  \caption{Cross-sectional shapes and maximum volumes}
  \renewcommand{\arraystretch}{1.2}
  \begin{tabular}{r|c}
    \hline
    Cross-sectional shape & Maximum volume \\
    \hline
    rhombus             & 0.243507 \\ %0.243507
    rectangle           & 0.243692 \\ %0.2436920418508385
    circle              & 0.278150 \\ %%0.2781499744348225
    arch (quad-Bézier)  & 0.294436  \\ %0.29443630190157716
    arch (cubic-Bézier) & 0.295448 \\ %0.2954475386100612
    arch (polyline with 1000 segments) & 0.295449 (0.174628 $^*$)\\ %0.2954494026680674
    \hline
  \end{tabular}
  
    * The value when the vertical dimension is set to 1.
   \label{tab:table}
\end{table}

\section{Conclusions}

In this paper, we focused on the fold curves, a critical element in determining the shape of a pillow box, and clarified their geometric constraints, demonstrating that a variety of shapes can be created within these constraints. We presented various design variations under the constraint of top-bottom symmetry, showcasing the diversity achievable in pillow box designs. Additionally, we demonstrated the feasibility of asymmetric top-bottom designs, expanding the scope of creativity in pillow box aesthetics.
This approach enhances the ability to freely control the shape of the curves, thereby broadening the range of design possibilities. As a result, it becomes possible to create more complex and aesthetically pleasing pillow boxes.

Another significant aspect of this research is the focus on optimizing the volume of pillow boxes. We addressed this by examining cases with circular, rectangular, and rhombus cross-sectional shapes, providing insights into determining the optimal dimensions for maximizing volume within design constraints, using both analytical and numerical methods. 
This approach allowed for understanding of the relationship between design parameters and the functional capacity of the pillow boxes.

Furthermore, for more general pillow box designs with an arch-shaped top surface, we applied quadratic B\'ezier curve, cubic B\'ezier curve and fine-polyline to the creation of folding curves. Through numerical calculations, we determined the shape of the curve that maximizes the volume. 
The results of this study confirm that the common pillow box shape with an arch-shaped top surface can be larger in volume than those with a circular, rectangular or rhombic cross-section. The shape of the curve that maximises the volume was also obtained by optimisation calculations. For the value of the maximum volume when the sheet is square, it was confirmed that the proposed method can yield a shape with a value 2.49\% larger than that calculated by Andrew Kepert \cite{web2}.

An approach to analytically determine the curve that maximises the volume is currently being worked on by Miyuki Koiso and the curve has been elucidated to be an elastic curve~\cite{koiso2023}. The specific type of elastic curve has not yet been clarified, but it is not expected to depart significantly from the shape and maximum volume values obtained in this paper.

In conclusion, this study not only deepens the understanding of pillow box design but also expands the possibilities of origami-based packaging design by discussing both aesthetic appeal and practical utility. 

Although we have focused on exploring the maximum volume of pillow boxes, if we consider the broader context of creating a three-dimensional object from two glued rectangular sheets of paper, the pillow box design might not always be the optimal choice. A cushion made of a square bag filled to the brim with cotton, or an inflated plastic balloon, will have a concave shape on all four sides while the centre is inflated. This shape cannot be easily made from paper and requires intricate folds.
Regarding the maximum volume achievable without limiting the design to pillow boxes, information can be found on the English Wikipedia under the {\it paper bag problem} or {\it teabag problem}. This problem investigates the maximum volume that can be achieved with a given amount of material, a fundamental question in packaging design. The maximum volume is proposed to be represented by an approximate formula.

\begin{equation}
V=w^3(h/(\pi w)-0.142(1-10^{(-h/w)})) 
\end{equation}

When the values of \( h = \sqrt{2} \) and \( w = 1 \) are substituted into the formula, the calculated volume is approximately 0.313629. The maximum volume of 0.295449 for the pillow box design demonstrated in this paper corresponds to about 94.2\% of this theoretical upper limit. This shows that the pillow box design is quite efficient and close to the theoretical limit, even though it can be realised with simple folds.

From a mathematical perspective, future work could include analytically determining the curve that maximizes volume. However, from a practical standpoint, verifying the strength and ease of assembly of these designs is crucial. Additionally, from a design aspect, loosening shape constraints and increasing the number of folds could introduce more variations in designs.

Given the significance of packaging as a practical application of origami, research should not be limited to the pillow box form but should encompass a broader range of shapes. This expanded focus could lead to the development of more innovative and versatile packaging solutions. 

\section*{Acknowledgement}
This work was supported by JST CREST Grant Number JPMJCR1911, Japan.

% reference section
\bibliographystyle{osmebibstyle}
\bibliography{osmerefs}

% author affiliations are appended at end of paper
\theaffiliations

\end{document}